\documentclass[prb,aps,twocolumn,showpacs,superscriptaddress]{revtex4-1}
\usepackage{amsmath}
\usepackage{graphicx, subfigure}
\usepackage{pslatex}
\usepackage{relsize}
\usepackage{bm}

\newcommand{\tiem}[1]{{\text{\tiny{#1}}}}

\makeatletter

\begin{document}

\title{Controlling the Polarity of the Transient Ferromagnetic-Like State  in Ferrimagnets}

\author{U.~Atxitia}
\affiliation{Department of Physics, University of York, York YO10 5DD, United Kingdom}
\affiliation{Departamento de Fisica de Materiales, Universidad del Pais Vasco, UPV/EHU, 20018 San Sebastian, Spain}

\author{J.~Barker}
\affiliation{Department of Physics, University of York, York YO10 5DD, United Kingdom}

\author{R. W.~Chantrell}
\affiliation{Department of Physics, University of York, York YO10 5DD, United Kingdom}

\author{O.~Chubykalo-Fesenko}
\affiliation{Instituto de Ciencia de Materiales de Madrid, CSIC, Cantoblanco, 28049 Madrid, Spain}

\begin{abstract}
  It was recently observed that the two antiferromagnetically coupled
  sublattices of a rare earth-transition metal ferrimagnet can temporarily
  align ferromagnetically during femtosecond laser heating, but always with the
  transition metal aligning in the rare earth direction. This behavior has
  been attributed to the slower magnetization dynamics of the rare earth sublattice.
  The aim of this work was to assess how the difference in the speed of the
  transition metal and rare earth dynamics affects the formation of the
  transient ferromagnetic-like state and consequently controls its formation.
  Our investigation was performed using extensive atomistic spin simulations
  and analytic micromagnetic theory of ferrimagnets, with analysis of a large
  area of parameter space such as initial temperature, Gd concentration
  and laser fluence.
  Surprisingly, we found that at high temperatures, close to the Curie point,
  the rare earth dynamics become faster than those of the transition metal.
  Subsequently we show that the transient state can be formed with the opposite polarity, where
  the rare earth aligns in the transition metal direction.
  Our findings shed light on the complex behavior of this class of
  ferrimagnetic materials and highlight an important feature which must be
  considered, or even exploited, if these materials are to be used in ultrafast
  magnetic devices.
\end{abstract}


\maketitle

\section{Introduction}

Unexpected states in solid state physics attract a lot of attention from
a fundamental point of view. A recent example is the transient
ferromagnetic-like state (TFLS) in ferrimagnetic GdFeCo.
\cite{RaduNature2011,OstlerNatureComm2012,KirilyukPhysRep2013,LeGuyaderAPL12}
 Element specific femtosecond resolution X-ray magnetic circular
 dichroism\cite{SiegmannBook2006} has shown that following the application of
 a femtosecond laser pulse, the FeCo sublattice magnetization points in the
 same direction as the Gd sublattice magnetization. This transient state lasts
 for just a few picoseconds, but the alignment against the strong
 antiferromagnetic coupling between the sublattices is very stable, even
 against large opposing magnetic fields.\cite{OstlerNatureComm2012}  More
 recent work has shown a similar behavior in TbFe alloys, \cite{KhorsandPRL13}
 suggesting this state may exist in a whole class of materials.
In this paper we study the TFLS in more depth, especially how the difference in
relaxation times of the transition metal (TM) and rare earth (RE) sublattices
leads to its formation.
We find that the TFLS can also form with the alignment of the rare earth
towards the transition metal sublattice, the opposite polarity to that which has
been found previously.
This change in polarity comes about due to a difference in the temperature
dependence of the characteristic relaxation times of the two sublattices,
leading to a crossover behavior, making the rare earth response faster than that
of the transition metal. This is a stark contrast to the commonly proposed
relation that the relaxation time of a magnetic species is proportional to the
ratio of its atomic magnetic moment $\mu_0$ and the intrinsic Gilbert damping
$\lambda$, \cite{KazantsevaEPL2008,MentinkPRL2012} a relation which would lead to
the false conclusion that the rare earth is always slower that the transition
metal.
Understanding this difference is important if these materials are to be used
for ultrafast magnetic devices, especially where large, dynamic temperature
ranges are used.

 Several theoretical descriptions have been proposed to describe the physical
 mechanisms underlying the TFLS.
 \cite{MentinkPRL2012,KoopmansPRB13,AtxitiaPRB2012,JoeSREP2013}  All of them
 suggest that the TFLS is  formed by an exchange of angular momentum between TM
 and RE magnetic lattices driven by the antiferromagnetic exchange coupling.
 The magnetization transfer leads to the TFLS in a non-equilibrium situation
 between both magnetic sublattices. The heating produced by the laser pulse
 causes the TM magnetization to nearly vanish, while that of the RE remains
 finite.
 Such a behavior is only possible if the demagnetization times (energy input
 absorption rate) of the two sublattices are sufficiently different. Therefore
 it is necessary to establish a theoretical approach to estimate the
 relationship between the relaxation times of each sublattice in ferrimagnets
 which would allow the engineering of generic ferrimagnetic materials, such as
 synthetic ferrimagnets, \cite{EvansArxiv2013} where this phenomenon could
 occur.

 In pure ferromagnets the ratio between the magnetic moment and the Curie
 temperature, $\mu_{0}/T_c$ (or more precisely \cite{AtxitiaQ} $\mu_{0}/\lambda
 T_c$)
  has been shown to be a good approximation of the \emph{speed} of
  demagnetization, allowing one to classify ferromagnetic materials into two types, fast and slow. \cite{KoopmansNatMat10}  Using this simplistic argument
  in TM-RE alloys such as TbFe, \cite{KhorsandPRL13} TbCo,\cite{AlebrandAPL12}
  or  GdFeCo \cite{OstlerNatureComm2012} and assuming a shared  $T_c$,
  \cite{OstlerPRB2011} transition metal demagnetization times
 are approximately $\mu_{\tiem{RE}}/\mu_{\tiem{TM}}\approx 4-5$ times faster
 than those of the rare earth. This simple relationship has been shown to work well for
 temperatures far below the Curie point. \cite{KoopmansNatMat10} This
 reasoning restricts the formation of the TFLS to a polarity where the TM
 magnetization will always reverse to align in the RE direction.

The importance of the initial temperature ($T_0$) on the demagnetization
behavior was shown in recent experiments on ferromagnetic Ni thin films where
the initial temperature before laser heating was systematically
varied.\cite{KoopmansPRX2012}  It was demonstrated that as $T_0$ increases the
observed \emph{fast} demagnetization becomes a well-defined two step process. Namely, an initial
sub-ps demagnetization followed by
a much slower demagnetization process of the order of several ps.
Similar results  where also obtained in Ni \cite{AtxitiaPRB10} and
FePt thin films\cite{Mendilarxiv2013} keeping the initial temperature fixed but increasing the laser fluency, and
thus reaching higher temperatures. The observed slowing down of the
magnetization response was associated to the increasing role played by the spin
fluctuations for temperatures approaching $T_c$.  In rare earth-transition
metal alloys, similar experiments have been recently performed in situations
where, although the initial temperature was varied, it did not approach $T_c$
closely enough to obtain clear evidence of the demagnetization
behavior.\cite{MedapalliEPJB2013,LopezfloresArxiv2013} From general phase
transition theory \cite{RushbrookeBook} one may expect that approaching $T_c$,
both sublattices would experience critical slowing down. However it is not
clear if the critical slowing down is equal for both sublattices and whether
the non-equivalence criterion,  $\mu_{\tiem{RE}}/\mu_{\tiem{TM}}$, still holds
close to $T_c$. To answer these questions, we theoretically investigate the
dynamical response of TM-RE alloys to femtosecond heating as a function of the
initial temperature and the TM-RE composition, using the prototypical example
of GdFeCo.

We use the Landau-Lifshitz-Bloch (LLB) equation for multi-sublattice magnets
\cite{AtxitiaPRB2012LLBferrim} to provide the quantitative calculation of the
magnetization relaxation times for each lattice, including those close to $T_c$
in GdFeCo. These calculations are complemented by large scale computer
simulations based on the stochastic Landau-Lifshitz-Gilbert (LLG) equation of
motion for an atomistic spin model.
These simulations confirm the proposed scenario for the formation of the TFLS
based on the angular momentum transfer between the two
sublattices.\cite{JoeSREP2013,AtxitiaPRB2012} Importantly, we derive
theoretically the necessary conditions for the formation of the TFLS with a
reversed polarity and confirm the criteria by means of atomistic spin dynamic
simulations. The paper is organized as follows. We first outline the basis of
the atomistic spin model followed by a consideration of the sublattice
relaxation times based on  the ferrimagnetic Landau-Lifshitz-Bloch equation
and importantly link this to the equilibrium magnetization curves. This leads to
a criterion for equivalent sublattice relaxation times, which defines a region
separating opposite polarities of TFLS; a result supported by the large-scale
atomistic model simulations.

\section{Theoretical description of ultrafast magnetization dynamics in two-sublattice ferrimagnets.}
\subsection{Atomistic spin model for G\lowercase{d}F\lowercase{e}C\lowercase{o} alloys}

The energetics of the ferrimagnetic system are described by the classical Heisenberg Hamiltonian
\begin{equation}
\mathcal{H} =
-  \tfrac{1}{2}\sum_{\langle ij \rangle}J_{ij}\mathbf{S}_{i}\cdot \mathbf{S}_j - \sum_{i} d_{z,i} (S_{z,i})^{2}
\label{eq:binaryalloyGeneralHamiltonian}
\end{equation}
where $\langle ij \rangle$ indicates that the sum is limited to nearest
neighbor pairs with $\left|\mathbf{S}_{i}\right|=\boldsymbol{\mu}_i/\mu_i$,
$\boldsymbol{\mu}_i$ representing the atomic magnetic moment.  We use a random
lattice model to represent the amorphous property of GdFeCo (see
Refs.~\onlinecite{OstlerNatureComm2012} and \onlinecite{JoeSREP2013}). We
assume a simple cubic lattice of FeCo moments and randomly substitute sites
with Gd moments until the desired Gd concentration, $x$, is achieved, giving
a system of Gd$_{x}$(FeCo)$_{1-x}$.  The FeCo sublattice has a moment
$\mu_{\tiem{FeCo}}=2.217 \mu_{\mathrm{B}}$ representing the $3d$ itinerant
electrons. \cite{BozorthBook2003}  The use of a common sublattice to model
FeCo is justified since the Fe and Co moments are delocalized in nature and are
parallel to one another up to the Curie temperature. Moreover, the amount of Co
used in the GdFeCo alloys studied in experiments is small. In our model the
Gd sublattice is attributed a moment of $\mu_{\tiem{Gd}}=7.63 \mu_{\mathrm{B}}$
which takes into account the contribution of the half-filled localized $4f$
shell ($7\mu_B$) and itinerant $5d$ electrons spin
($0.63\mu_B$)\cite{JensenBook1991} ($\mu_{\mathrm{B}}$ is the Bohr magneton in
both cases). The values of exchange energy, $J_{ij}$, we use give good
agreement to the experimental data of GdFeCo for the Curie temperature, $T_c$, and
compensation point, $T_M$, where $M_{\mathrm{FeCo}}=M_{\mathrm{Gd}}$.  Here,
$M_{\mathrm{FeCo}}(T)=(1-x)\mu_{\tiem{FeCo}} \langle
S_{\mathrm{FeCo}}\rangle(T)$ and  $M_{\mathrm{Gd}}(T)=x\mu_{\tiem{Gd}} \langle
S_{\mathrm{Gd}}\rangle(T)$ are the macroscopic magnetization of the FeCo and Gd
sublattices, respectively.

The thermodynamic average the spin fluctuations of each sublattice, $ \langle
S_{\mathrm{Gd}}\rangle(T)$ and  $ \langle S_{\mathrm{FeCo}}\rangle(T)$,  can be
calculated using either the mean field approximation or using computational
models.  \cite{OstlerPRB2011}  In the atomistic spin model we will use computer
simulations with the values for exchange energy  $J_{\tiem{RR}}=
2.3970\times10^{-21}$J and $J_{\tiem{TT}}=
1.0067\times10^{-20}$J (ferromagnetic coupling) and
  $J_{\tiem{RT}}=-2.7805\times10^{-21}$J (antiferromagnetic coupling).  Typical
  Gd concentrations used in recent experiments of ultrafast dynamics range from
  $10$ to $30 \%$. \cite{KirilyukPhysRep2013} Hence, these alloys behave as an
  ``effective'' ferromagnet, FeCo, defined mainly by the value of the
  intra-lattice exchange integral, $J_{\tiem{TT}}$, and the Gd spins whose
  dynamics are mainly defined by the  antiferromagnetic coupling to the FeCo
  spins, defined by the exchange integral $J_{\tiem{TR}}$. The effect of Gd-Gd
  exchange interactions in these Gd concentrations ($10-30\%$) is relatively
  small in comparison to other exchange interactions present in the system.  We
  note that although in this work we are studying GdFeCo alloys in particular,
  the model described above is general and can be adapted to any other rare earth
  - transition metal alloy. The second term in the Hamiltonian represents
  a uniaxial anisotropy energy which is consistent with the observed out of
  plane magnetization of these alloys. We use a value of $d_{z} = 8.07246\times
  10^{-24}$ J for both FeCo and Gd sublattices.\cite{OstlerPRB2011}

 The dynamics of the atomistic spins interacting with a heat bath are
 described by the coupled stochastic Landau-Lifshitz-Gilbert (LLG) equations of
 motion
 \begin{equation}
   \frac{\mathrm{d} \mathbf{S}_{i}}{\mathrm{d} t} =- \frac{\gamma_{i}}{(1+\lambda_{i}^{2})} \left[\mathbf{S}_{i}\times\mathbf{H}_{i,\mathrm{eff}} + \lambda_{i} \mathbf{S}_{i}\times(\mathbf{S}_{i}\times\mathbf{H}_{i,\mathrm{eff}})\right]
\label{eq:stochasticLLG}
 \end{equation}
 where $\gamma_{i}$ is the gyromagnetic ratio, $\lambda_{i}$ is the coupling
 strength of the spin $i$ to the heat bath and the effective field of a spin on
 lattice site $i$ is
 \begin{equation}
 \mathbf{H}_{i,\mathrm{eff}}
   = -\frac{1}{\mu_{i}} \frac{\partial \mathcal{H}}{\partial \mathbf{S}_{i}
 } + \boldsymbol{\zeta}_{i}.
 \end{equation}
 The stochastic fields,
 $\boldsymbol{\zeta}_{i}$, represent the thermal fluctuations. Although one can
 in principle consider colored thermal noise, \cite{AtxitiaPRL2009} here we
 assume that it is uncorrelated in space and time, \emph{i.e.} the white noise
 approximation. The first and second moments of the bath variable are written
 as:
 \begin{equation}
 \langle \zeta^{k}_i \rangle
 = 0, \quad
 \langle \zeta^k_i (t)\zeta^l_j (t')\rangle =2 \lambda_{i} \frac{\mu_i }{\gamma_{i}}k_{\mathrm{B}}T\delta_{ij}\delta_{kl}\delta(t-t'),
 \end{equation}
 where $i,j$ denotes lattice sites and $k,l$ are the Cartesian components,
 $k_B$ is the Boltzmann constant, and $T$ is the heat-bath temperature.  We
 couple the spin system to a heat bath representing the electronic system  and
 defined by the temperature $T_e$.  This acts as the origin of the spin-flip
 processes and is where energy enters the system from the laser heating.
 \cite{KazantsevaPRB2008} The coupling constant $\lambda_i$ essentially
 represents the energy transfer rate between the spin and heat bath.  Here each
 sublattice is separately coupled to the same heat bath, although they could be
 coupled to different heath baths, such as electron and phonon heat baths as shown in
 Ref. \onlinecite{WienholdtPRB2013} or the same sublattice can be
 coupled simultaneously to the two baths. \cite{SultanPRB2012} As a consequence
 of the different magnetization response of each sublattice, given by the intrinsic
 parameters, one sublattice can heat faster (``hot'') than the other
 (``cold''). The absorbed energy is afterwards dissipated in both the heat bath
 (fluctuation-dissipation) and to the other sublattice (and eventually  to the
 heat bath) through the atomic scale exchange coupling. \cite{EllisPRB2012} The
 transient ferromagnetic like state in GdFeCo is formed under non-equilibrium
 situations where the latter mechanism dominates. \cite{AtxitiaPRB2012}

Atomistic spin models of this type have become a \emph{de facto} tool in the
investigation of ultrafast magnetism due to their good quantitative description
of the temperature dependent magnetic properties, as well as the timescales of
dynamical phenomenon such as the TFLS.  In principle one can include many
complexities into such a model, for example different values of $\gamma_{i}$
and $\lambda_{i}$. Such scenarios may be closer to the physical reality,
however to clearly explain the most basic relationships behind the dynamics
which cause the formation of the TFLS we assume $\gamma_{\mathrm{FeCo}}
= \gamma_{\mathrm{Gd}} = \gamma_{e}$ and $\lambda_{\mathrm{FeCo}}
= \lambda_{\mathrm{Gd}} = 0.01$.  These values are the same as those used in
previous works and describe well the ultrafast demagnetization time scales
found experimentally. \cite{OstlerNatureComm2012}

\subsection{Formation of the transient ferromagnetic like state}\label{sec:formation}

The role of the magnetization compensation point in the laser induced
magnetization switching and the TFLS formation is still a subject of
controversy in the literature. \cite{StanciuPRL2007, MekomenPRL2011,
WienholdtPRL12, AlebrandPRB2012,
HassdenteufelAdvanceMaterials13,SchlickeiserPRB2012, JoeSREP2013}  Recently we
showed that a minimum amount of laser energy input  of the order of the energy
gap between  the two ferrimagnetic  modes, \emph{i.e.} $\hbar \Delta \omega
\sim \hbar \omega_{\mathrm{ex}}\sim  J_{\tiem{RT}} \left(M_{\mathrm{FeCo}}
(T_0)-M_{\mathrm{Gd}}(T_0) \right)$, is required for the formation of the
TFLS,\cite{JoeSREP2013} where  $\omega_{\mathrm{ex}}$ is the frequency of the
so-called  antiferromagnetic coherent exchange mode.  The disordered nature of
the GdFeCo spin lattice leads an effect that the most efficient energy
transfer (the smallest energy gap) does not correspond exactly to the coherent
mode with $k=0$ but to non-zero $k$ value, related to the characteristic
length of the Gd spins cluster size.\cite{JoeSREP2013}

This slightly shifts the minimum energy for the TFLS formation
with respect to the magnetization compensation point, where
$ M_{\mathrm{FeCo}} (T_0)\approx M_{\mathrm{Gd}}(T_0)$ but the minimum energy
still lies close to it. The condition for the minimum energy gap can be
fulfilled in two temperature regions (i) close to the $T_M$ where
$M_{\mathrm{FeCo}} (T_M)\approx M_{\mathrm{Gd}}(T_M)$, (with
$ M_{\mathrm{FeCo}} (T_M) \neq 0, M_{\mathrm{Gd}}(T_M)\neq 0$), and (ii)
approaching the Curie temperature $T_c$, where both $ M_{\mathrm{FeCo}} (T_M)$
and $M_{\mathrm{Gd}}(T_M)$ are close to zero. The temperature dependence of
the  ferrimagnetic exchange modes depends on the concentration of Gd ($x$), the
initial temperature ($T_0$) and the inter-lattice exchange coupling
($J_{\tiem{RT}}$). \cite{SchlickeiserPRB2012} Among these parameters  $x$ and
$T_0$ can be changed experimentally however $J_{\tiem{RT}}$ is very difficult
to change in a GdFeCo alloy. We therefore focus on the fundamental properties of
the TFLS formation by varying the concentration and initial temperature.
\begin{figure}[tbc!]
\begin{centering}
\includegraphics[width=0.5\textwidth]{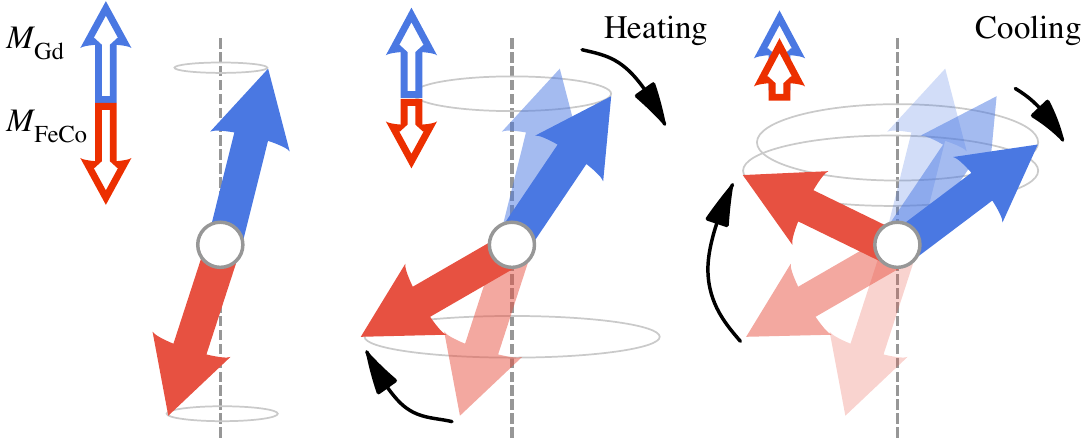}
\end{centering}
\caption{
Schematic description of the transition into the transient ferromagnetic-like
state (TFLS) with rare earth (RE) polarity. Initially, at equilibrium, there is
little excitation of high energy antiferromagnetic exchange spin oscillations.
During the heating phase these are excited and the TFLS is formed with RE
polarity since the RE dynamics are faster.
}
\label{fig:Schematic}
\end{figure}

Figure \ref{fig:Schematic} shows a schematic representation of the transient
ferromagnetic like state formation.  As explained in detail in
Ref.~\onlinecite{JoeSREP2013}, three well defined regions can be distinguished
(i)
at equilibrium before laser heating, the thermal excitation of the modes does
not lead to any net spin transfer between sublattices,
(ii)
during laser heating, the spin wave modes of the system are excited out of
equilibrium,  leading to an efficient transfer of spin angular momentum between
sublattices. \cite{JoeSREP2013} (iii) This process results in a precessional reversal
path for the magnetization at the nanoscale \cite{AtxitiaPRB2012}.  Around
$T_M$ the energy efficiency of this mechanism is large since the relevant modes
are lower in energy ($\hbar\omega_{\textrm{ex}}(T_M)\approx 0$) than when the
system is far from $T_M$.  \cite{JoeSREP2013}
Nevertheless, this does not restrict the appearance of the TFLS to the existence of any
compensation point. Indeed the TFLS has been shown above the compensation
point both in spin models and experimentally. \cite{OstlerNatureComm2012}

Apart from the excitation of those modes, before the TFLS can form, a full
macroscopic demagnetization of the magnetic sublattice which switches (hot) is
needed, \cite{OstlerNatureComm2012} whereas the other remains finite (cold).
Hence, the heating produced by the laser pulse into one of the sublattices should
have the same time scale as the magnetization relaxation time $\tau_{\nu,\|}$,
with $\nu=$RE,TM. Thus, the estimation of $\tau_{\nu,\|}$ is of fundamental
importance.
\begin{figure}[tbc!]
\begin{centering}
\includegraphics[width=0.5\textwidth]{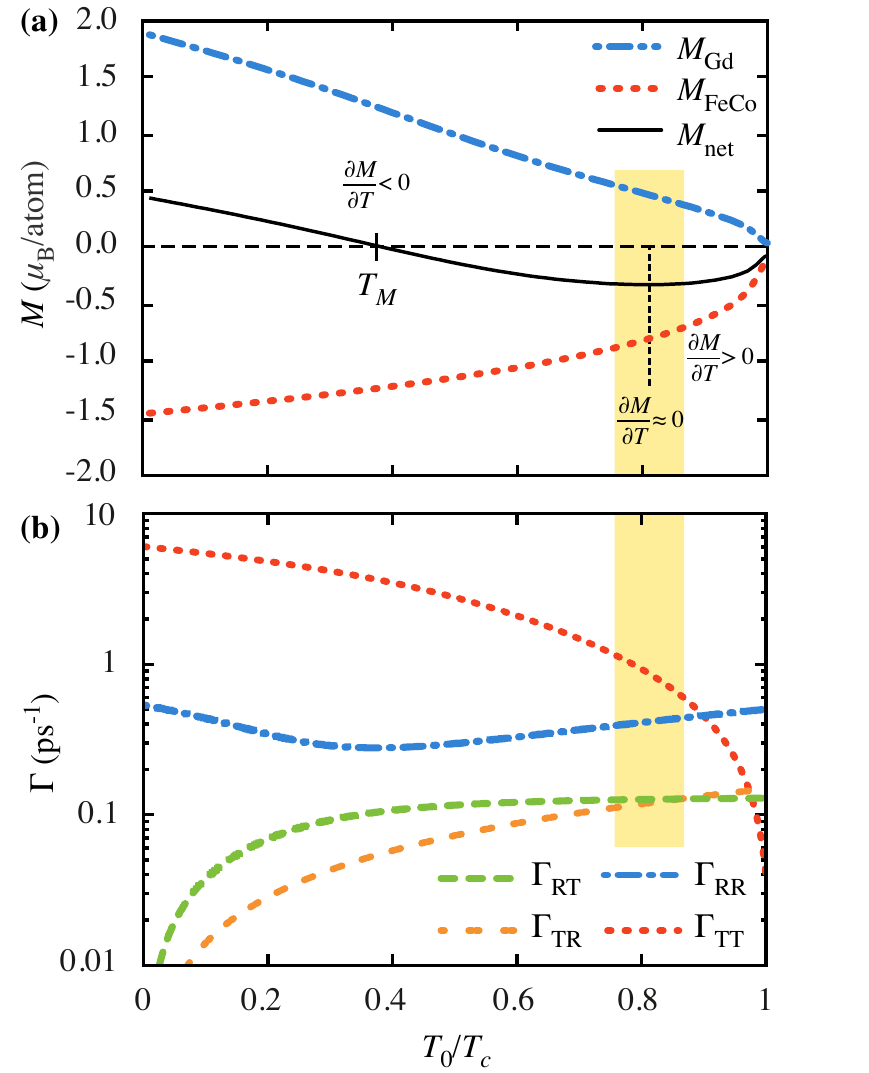}
\end{centering}
\caption{
(a) Equilibrium magnetization of each sublattice as a function of reduced
temperature $T_0/T_c$ ($T_c=800$ K).  The yellow shaded region corresponds to
$\partial M_{\textrm{net}} /\partial T=0$ and separates two regions, $\partial
M_{\mathrm{FeCo}}/\partial T<\partial M_{\mathrm{Gd}}/\partial T$  and
$\partial M_{\mathrm{Gd}}/\partial T>\partial M_{\mathrm{FeCo}}/\partial T$.
(b) Relaxation rate parameters defining the Landau-Lifshitz-Bloch dynamical
equation as a function of  $T_0/T_c$.
}
\label{fig:Fig1}
\end{figure}

\subsection{Estimation of the temperature dependent magnetization relaxation time in ferromagnets}

 The demagnetization behavior of even simple ferromagnets, after a femtosecond
 timescale laser pulse, is still a subject of intensive research.
\cite{MannPRX2012,WietstrukPRL2011,BattiatoPRL2010,SchellekensPRL2013,AtxitiaQ,ManchonPRB2012,MuellerPRL2013}
 In ferrimagnets, there is a large interest on demagnetization rates of each
 species, following the use of X-ray magnetic circular dichroism
 \cite{SiegmannBook2006} to monitor the sub-picosecond dynamics of each sublattice
 magnetization individually. \cite{KirilyukPhysRep2013} Thermal effects
 dominate the magnetization dynamics during the heating produced by the laser.
 On this basis recent works \cite{KazantsevaEPL2008,MentinkPRL2012} suggested
 that the typical demagnetization time of the sublattice $\nu$ is given  by
 $\tau_{\nu} \sim \gamma_{\nu}\mu_{\nu}/( \lambda_{\nu} k_{\mathrm{B}} T_e)$,
 where $T_e$ refers to the dynamical electron temperature. This estimate is
 based on the assumption that the magnetic system behaves as a paramagnet
 before or instantaneously after the application of the heating.  Effectively
 this assumes that the dynamics is completely dominated by temperature and all
 spin correlations are lost, thus this time scale takes no account of the
 magnetic interactions. This assumption is extremely hard to justify in light
 of recent theories which suggest the spin correlations play an important role
 in forming the TFLS.\cite{JoeSREP2013} Nevertheless, for GdFeCo alloys, it has
 been found to give a good estimate of the relationship between the
 demagnetization times of FeCo and Gd sublattice when compared to experimental
 observations where
 $\tau_{\tiem{Gd}}/\tau_{\tiem{FeCo}}\approx\mu_{\tiem{Gd}}/\mu_{\tiem{FeCo}}
 \approx 3.5$ (a  $\lambda_{\tiem{Gd}}=\lambda_{\tiem{FeCo}}$ were assumed).
 \cite{RaduNature2011} This coincidence has caused some confusion and restricts
 research on this new phenomenon to materials with very different atomic
 magnetic moments.

 More sophisticated approaches such as the Landau-Lifshitz-Bloch (LLB)
 \cite{AtxitiaPRB10,AtxitiaQ} and microscopic three temperature model (M3TM)
 \cite{KoopmansNatMat10} models take into account the spin exchange
 interactions within the mean field approximation. The M3TM model also within the
 MFA (Weiss) shows that the timescale is essentially defined by $\mu_0/T_c$.
 Within the LLB model a similar estimate can been found.\cite{AtxitiaQ} In
 particular, for ferromagnets, the relaxation time has been shown to be defined
 as $\tau_{\|} \sim \widetilde{\chi}_{\|}(T)/ \lambda$, where
 $\widetilde{\chi}_{\|}=(\partial m/ \partial H)_{H\rightarrow \infty}$ is the
 longitudinal susceptibility and $m$ is the thermally averaged spin
 polarization. This depends on: (i) the particular dissipation mechanism
 represented by the coupling parameter $\lambda$, and (ii) the longitudinal
 susceptibility  $\widetilde{\chi}_{\|}(T)$.  In the LLB model $\lambda$ is
 related to the spin-flip scattering probabilities, and taken as
 a parameter \cite{GaraninPhysicaA91,UnpublishedPablo} while within the M3TM
 model it is related to electron-phonon Elliot-Yafet scattering mechanism.
 \cite{SchellekensPRL2013}

 In the MFA the longitudinal susceptibility can be calculated using the
 equilibrium magnetization curve given by the Curie-Weiss equation
 $m=\mathcal{L}(mJ_{0}/k_{\mathrm{B}} T)$, where $\mathcal{L}$ stands for the
 Langevin function in the classical case, $\mathcal{L}(x)=\coth(x)-1/x$ and
 $J_0=\sum_{j}J_{0j}=3k_BT_c$. It reads
 \begin{equation}
 \widetilde{\chi}_{\|}(T)=
 \frac{\mu_0}{J_0} \frac{\beta J_0 \mathcal{L}'}{1-\beta J_0\mathcal{L}'}\approx
 \left\{ \begin{array}{ccc}
  c_0\frac{\mu_0}{T_c},& & T \ll T_{c}\\
& &\\
\frac{\mu_0}{2} \frac{T_c}{T-T_c}, &
& T\lesssim T_{c}
\end{array}
\right.
\label{eq:longitudinalsusferro}
 \end{equation}
 where $\mathcal{L}'=\mathrm{d} \mathcal{L}(x)/\mathrm{d} x$.  At low
 temperature $T \ll T_{c}$, the function $c_0=\beta J_0 \mathcal{L}'/(1-\beta
 J_0\mathcal{L}')$  in Eq. \eqref{eq:longitudinalsusferro} is almost constant
 with increasing temperature. We note that we have used the notation $c_0$,
 similar to that in Ref. \onlinecite{KoopmansPRL05}, where it was proposed for
 the first time. Thus, the magnetization relaxation time in this region is
 mainly determined by $\tau_{\|}  \sim \mu_0/\lambda T_c$ (or more exactly,
 $\tau_{\|}  \sim c_0 \mu_0/(\gamma \alpha_{\|}T_c)$. Note that the dependence
 on $T_c$ comes from its linear relationship with the exchange parameter $J_0$,
 that sets the energy scale of the spin fluctuations. Within the classical spin model used here, the equilibrium
 magnetization at low temperatures  decreases linearly with $T$, thus
 $|\partial m/\partial T|\approx\textrm{const}$. We note that for a quantum spin approach, at
 low temperature the magnetization follows the well-known Bloch $T^{3/2}$ law, and therefore, one will obtain  $|\partial m/\partial T|\approx T^{1/2}$.

For temperatures close to $T_c$ the situation is different and the
magnetization scales as $m^2\simeq 5/3(1-T/T_c)$, which leads to
$\widetilde{\chi}_{\|}\propto (\partial m/\partial T)^2 (T/T_c)\approx
(\partial m/\partial T)^2\propto (1-T/T_c)^{-1}$, describing the well-known
magnetization critical slowing down close to the Curie temperature.  The
behavior of the $\partial m/\partial T$ can be interpolated for intermediate
temperature regions.  Thus the derivative $\partial m/\partial T$ gives
a reasonable qualitative description about the demagnetization speed.  This is
not surprising as the physical origin of the equilibrium magnetization
temperature dependence and the thermally induced longitudinal relaxation dynamics
is the same - the thermal excitation of the spin waves.

\subsection{Bloch relaxation of ferrimagnets as a function of temperature}
\begin{figure}[tbc!]
\begin{centering}
\includegraphics[]{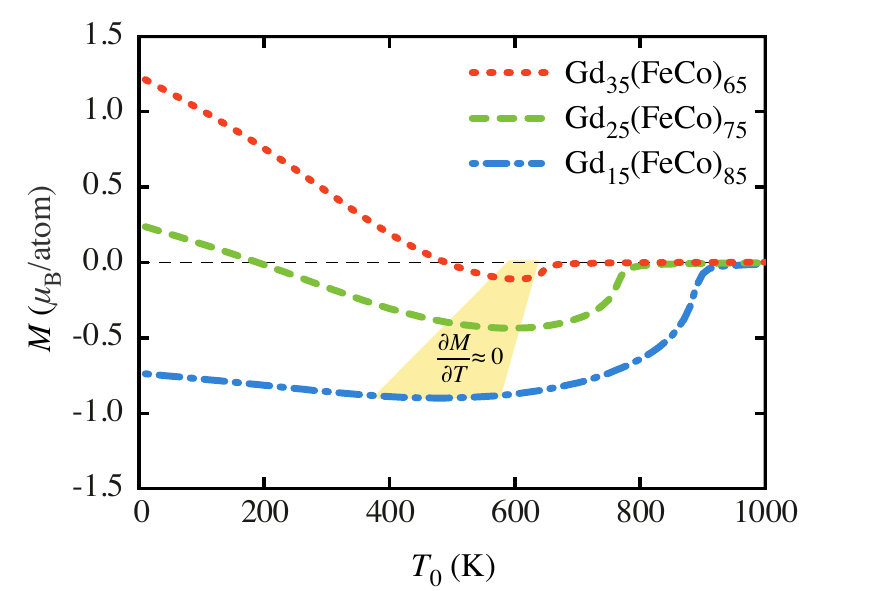}
\end{centering}
\caption{(Color online) Equilibrium magnetization of Gd$_x$(FeCo)$_{1-x}$
  alloys as a function of temperature for three Gd concentrations,
  $x_{\mathrm{Gd}}=0.15, 0.25,$ and $0.35$. The yellow shaded area represents the
  temperature region where $\partial M_{\mathrm{FeCoGd}}/\partial T\approx 0$.
}
\label{fig:equilibriumM}
\end{figure}
 In contrast to ferromagnets, where the net equilibrium magnetization $M(T)=M_s(T=0 \ \mathrm{K}) m(T)$
 always decreases with increasing temperature, in RE-TM ferrimagnetic alloys
 more possibilities exist. As an example to illustrate this, we consider the
 disordered alloy Gd$_{25}$(FeCo)$_{75}$ (with the parameters of Ref.
 \onlinecite{OstlerPRB2011}). Here, as shown in Fig~\ref{fig:Fig1}(a), the net
 magnetization goes through a transition at a ``magnetization derivative''
 compensation temperature $T_{\textrm{tr}}$  at $\partial
 M_{\mathrm{net}}/\partial T = 0$, where $M_{\mathrm{net}}=M_{\mathrm{Fe}}
 x_{{\mathrm{Fe}}}m_{\mathrm{Fe}}-M_{\mathrm{Gd}}
 x_{{\mathrm{Gd}}}m_{\mathrm{Gd}}$.  Because of the slow variation of the
 magnetization around $T_{\textrm{tr}}$ (see Fig.~\ref{fig:Fig1}(a)), it is
 most appropriate to define the temperatures around $T_{\textrm{tr}}$ as
 a region in which a transition between two different demagnetization rates
 occur. We will show that this is the area in which a transition occurs between
 the two polarizations of the transient state.  This is dependent on the alloy
 concentration as illustrated in Fig.~\ref{fig:equilibriumM}, which shows the
 equilibrium magnetization of Gd$_x$(FeCo)$_{1-x}$  alloys as a function of
 temperature for three Gd concentrations, $x_{\mathrm{Gd}}=0.15, 0.25,$ and
 $0.35$. The yellow shaded area represents the temperature region where
 $\partial M_{\mathrm{FeCoGd}}/\partial T\approx 0$.
\begin{figure*}[tbc!]
\begin{centering}
\includegraphics[width=0.98\textwidth]{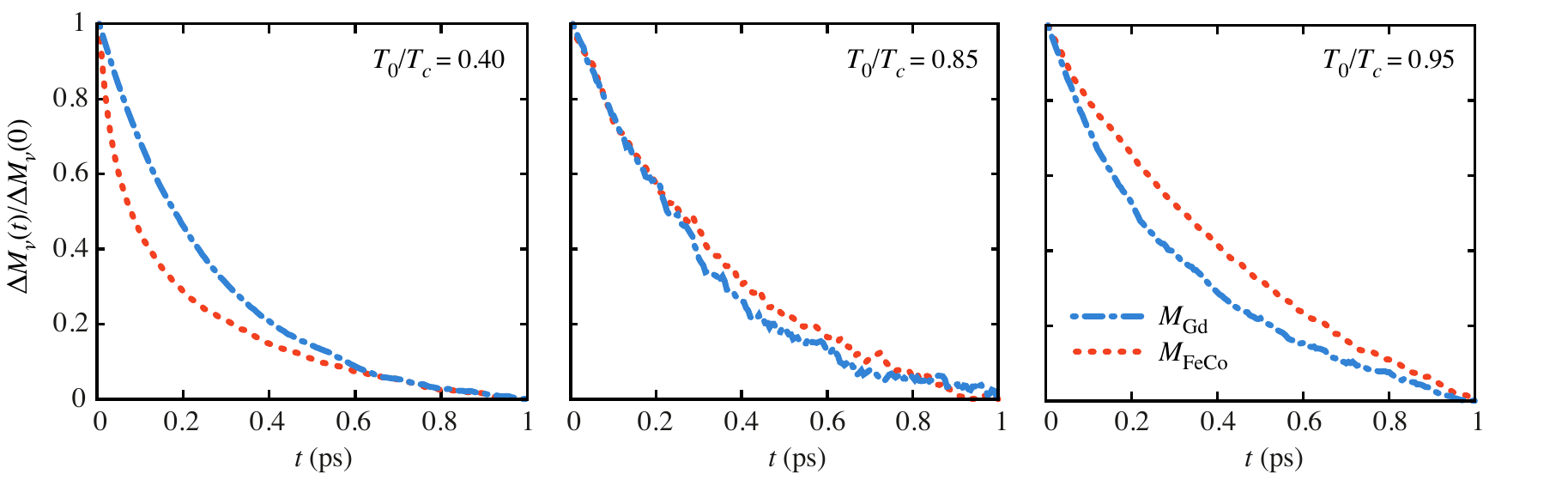}
\end{centering}
\caption{
  Dynamic relaxation after a step increase in temperature of the normalized magnetization of FeCo and Gd sublattices for three different temperatures as calculated from the atomistic approach. We use the normalization
magnetization, $\Delta M_{\nu}(t)/ \Delta M_{\nu}(0)$ rather than $M_{\nu}(t)$ to directly compare
the relative magnetization relaxation dynamics of each sublattice. Note the change in relative relaxation rates from 'slow' Gd at low temperatures to 'fast' Gd approaching $T_c$.
}
\label{fig:relaxation}
\end{figure*}

Subsequently, and still for our exemplar alloy system  Gd$_{25}$(FeCo)$_{75}$,
we can define three well-defined temperature regions in terms of partial
sublattice magnetization derivatives
 \begin{equation}
   \left|\frac{\partial M_{\mathrm{FeCo}}}{\partial T}\right|
   < \left|\frac{\partial M_{\mathrm{Gd}}}{\partial T}\right|
   \quad \mathrm{for} \quad T \lesssim \frac{3}{5}T_c \\
 \end{equation}
 where the FeCo sublattice magnetization responds faster to temperature changes
 than the Gd sublattice. Around the point of inflection we have
 \begin{equation}
   \left|\frac{\partial M_{\mathrm{FeCo}}}{\partial T}\right|
   \approx \left|\frac{\partial M_{\mathrm{Gd}}}{\partial T}\right|
   \quad \mathrm{for} \quad \frac{3}{5}T_c \lesssim T \lesssim \frac{4}{5}T_c \\
 \end{equation}
 and so both sublattices respond to temperature changes on a similar timescale.
 Finally for high temperatures close to $T_c$ we have
 \begin{equation}
   \left|\frac{\partial M_{\mathrm{FeCo}}}{\partial T}\right|
   > \left|\frac{\partial M_{\mathrm{Gd}}}{\partial T}\right|
   \quad \mathrm{for} \quad  T_c \gtrsim T \gtrsim \frac{4}{5}T_c \\
 \end{equation}
indicating a temperature region where the FeCo sublattice magnetization
responds slower than that of the Gd.

These criteria can be verified by comparing the quasi-equilibrium relaxation
dynamics of each sublattice magnetization using the many-spin stochastic LLG
equations \eqref{eq:stochasticLLG}, after a step like increase in temperature
of $\Delta T = T_{0}/10$, as shown in Fig.~\ref{fig:relaxation}.  We present
the results of our simulations in each regime, $T_0/T_c = 0.4,0.85$ and $0.95$
($T_c = 800$ K). These results confirm the existence of the three distinct
regions discussed previously.  Most importantly, at temperatures very close to
$T_c$ we do find that the Gd magnetization relaxes faster than that of the FeCo
sublattice, contrary to the common perception that $\mu_0/T_{c}$ or
$\mu_0/\lambda$ determine this
timescale.\cite{KoopmansNatMat10,RaduNature2011,MentinkPRL2012}

Therefore, our initial postulate that the relative speed of each magnetic
lattice can be deduced from the value of $\partial M/\partial T$ is supported
by the atomistic simulations.  This is an important result relating the
boundary between regimes of dynamic behavior to the relatively straightforward
measurement of a static magnetic property. If this finding is experimentally
confirmed, it will be very useful in the design of new materials showing novel
ultrafast switching properties. However, this criterion alone does not provide
a direct quantitative description of the relaxation behavior from the magnetic
properties of the system. In the following we investigate the magnetization
dynamics using an analytical approach based on the ferrimagnetic LLB equation.
This provides an important basis for the interpretation of the detailed dynamic
behavior calculated using the atomistic spin model.

\subsection{Relaxation times derived from the Landau-Lifshitz-Bloch equation}

Atomistic spin modeling has proved very adept at describing the non-equilibrium
spin dynamics of GdFeCo in direct comparison with those observed in
experiments.\cite{RaduNature2011,OstlerNatureComm2012,WienholdtPRL12} However,
the many body nature of this approach makes the physical interpretation of
results difficult. Moreover, \emph{a priori} predictions are not possible from
such an approach.  In contrast, the macroscopic Landau-Lifshitz-Bloch (LLB)
equation\cite{GaraninPRB1997}  provides analytical expressions for the
relaxation rates in terms of the magnetic properties of the system.  The LLB
equation describes the non-equilibrium average magnetization dynamics and is
derived starting from the Heisenberg Hamiltonian
\eqref{eq:binaryalloyGeneralHamiltonian} and the Fokker-Planck equation based
on the atomistic Landau-Lifshitz equation thus providing a macroscopic
description of the atomistic dynamics. \cite{GaraninPRB1997}  Hence, we will
use the LLB model to give clear evidence for the heuristic relation presented
previously between the relaxation time and the proposed criterion in terms of
$\partial M/\partial T$.

The ferrimagnetic LLB equation\cite{AtxitiaPRB2012LLBferrim} is written for the
reduced magnetization of each sublattice, $\langle \mathbf{S}_{\nu}\rangle
=\mathbf{m}_{\nu}=\mathbf{M}_{\nu}/M_{s,\nu}$, where ${\nu}$ denotes TM or RE
sublattice and $M_{s,\nu}$ is the saturation magnetization at $T=0$ K
\begin{eqnarray}
\frac{\mathrm{d}\mathbf{m}_{\nu}}{\mathrm{d} t}&=&
-|\gamma_{\nu}|\mathbf{m}_{\nu}\times
\Big[
\mathbf{H}_{\text{eff}}^{\nu} +
\frac{\alpha^{\nu}_{\perp}}{m^2_{\nu}}
\mathbf{m}_{\nu}
\times
\mathbf{H}_{\text{eff}}^{\nu}
\Big]+|\gamma_{\nu}|\alpha^{\nu}_{\|} H_{\|}^{\nu}\mathbf{m}_{\nu}
  \nonumber \\
& &
\label{eq:LLBferrimagFinalexpression1}
\end{eqnarray}
The first two terms describe the motion of the transverse magnetization
components, with a form form similar to the well-known
Landau-Lifshitz-Gilbert equation, and which are driven by the effective field
$\mathbf{H}_{\text{eff}}^{\nu}$ comprised of the usual anisotropy and applied
fields. These two terms do not affect the pure longitudinal motion, the
relaxation of which we are interested in. This Bloch-type (longitudinal)
relaxation is described by the third term in Eq.
\eqref{eq:LLBferrimagFinalexpression1}.  Therefore we will neglect the
transverse components in the following to reduce the number of parameters in
Eq. \eqref{eq:LLBferrimagFinalexpression1}. Given the dominance of the
Bloch-type relaxation this is a reasonable assumption that results in useful
analytic expressions. Thus, the longitudinal Bloch-type relaxation is defined
by
\begin{equation}
\frac{\mathrm{d} m_{\nu}}{\mathrm{d} t}= |\gamma_{\nu}|\alpha^{\nu}_{\|} H_{\|}^{\nu} m_{\nu}.
\end{equation}
To further simplify the interpretation of the relaxation rates defining this equation and
since we are interested in the relative rather the absolute relaxation times we
can assume $\gamma_{\tiem{TM}}=\gamma_{\tiem{RE}}$  and
$\lambda_{\|}^{\tiem{T}}=\lambda_{\|}^{\tiem{R}}$.
 The longitudinal  field $H_{\|}^{\nu}$ which defines the different dynamics of each sublattice reads
\begin{eqnarray}
H_{\|}^{\nu} & = &
\left[\frac{1}{2\Lambda_{\nu\nu}}\left(\frac{m_{\nu}^{2}}{m_{e,\nu}^{2}}-1\right)-
\frac{1}{2\Lambda_{\nu\kappa}}\left(\frac{m_{\kappa}^2}{m_{e,\kappa}^2}-1\right)\right].
\label{eq:longitudinalfield}
 \end{eqnarray}
The subscript $e$ denotes the equilibrium values. The field in
Eq.~\eqref{eq:longitudinalfield} is comprised of the difference between the
relaxation of sublattice $\nu$ to its own equilibrium value and to the
equilibrium value of the $\kappa$ sublattice. The relaxation parameters are
defined in Table \ref{table_parameters}. They are expressed in term of
a physically measurable longitudinal susceptibilities. The latter can be also
evaluated in the MFA approximation as
\begin{widetext}
\begin{equation}
\chi^{\nu}_{\|}=\frac{\mu_{\kappa}}{|J_{0,\kappa\nu}|}\frac{-|J_{0,\nu\kappa}||J_{0,\kappa\nu}|\beta^{2}\mathcal{L}_{\kappa}^{'}(.)\mathcal{L}_{\nu}^{'}(.)
+(\mu_{\nu}/\mu_{\kappa})|J_{0,\kappa\nu}|\beta \mathcal{L}_{\nu}^{'}(.)(1-J_{0,\kappa}\beta \mathcal{L}_{\kappa}^{'}(.))}{(1-J_{0,\kappa}\beta \mathcal{L}_{\kappa}^{'}(.))(1-J_{0,\nu}\beta \mathcal{L}_{\nu}^{'}(.))-|J_{0,\nu\kappa}||J_{0,\kappa\nu}|\beta^{2}\mathcal{L}_{\kappa}^{'}(.)\mathcal{L}_{\nu}^{'}(.)}
\label{eq:susceptibilities}
\end{equation}
\end{widetext}
where the parameter definitions  are given in table \ref{table_parameters}. We note that the
relaxation parameters in Eq. \eqref{eq:susceptibilities} depend on the exchange
and the atomic magnetic moments of both sublattices.

\begin{table}[]
	\begin{center}
\begin{tabular}{ccc}
\hline
\hline
Parameter & Expression & Description   \\ \hline
$\Lambda_{\nu\kappa}$ & $|J_{0,\nu\kappa}|/\mu_{\nu}m_{e,\nu}/$  &  Relaxation parameter   \\
$\Lambda_{\nu\nu}$ & $\widetilde{\chi}^{\nu}_{\|}/\left(1-|J_{0,\nu\kappa}|\widetilde{\chi}^{\kappa}_{\|}/\mu_{\nu}
\right)$    & Relaxation parameter   \\
$\alpha_{\Vert}^{\nu}$ & $2\lambda_{\nu}/\beta \widetilde{J}_{0,\nu}$ & Longitudinal damping   \\
$\alpha_{\bot}^{\nu}$ & $\lambda_{\nu}(1-1/\beta \widetilde{J}_{0,\nu})$  &  Perpendicular damping \\
$\widetilde{J}_{0,\nu}$& $J_{0,\nu}+|J_{0,\nu\kappa}|m_{e,\kappa}/m_{e,\nu}$ & MFA exchange   \\
$J_{0,\nu}$& $x_{\nu}zJ_{\nu\nu}$ & Intra-lattice exchange   \\
$J_{0,\nu\kappa}$& $x_{\kappa}zJ_{\nu\kappa}$ & Inter-lattice exchange   \\
\hline \hline
\end{tabular}
\caption{Explicit expressions of the parameters entering  Eqs.
  \eqref{eq:LLBferrimagFinalexpression1} and  \eqref{eq:longitudinalfield}.
  $\widetilde{\chi}^{\nu}_{\|}=\left(\partial m_{\nu}/\partial
  H\right)_{H\rightarrow 0}$, and $m_{e,\nu}$ are calculated using the MFA
  \cite{AtxitiaPRB2012LLBferrim} $\beta=1/k_{\mathrm{B}} T$.
 $x_{\nu}$ is the concentration of the specie $\nu$ and $z$ the number of nearest neighbours.
}
\label{table_parameters}
\end{center}
\end{table}

To calculate the relaxation rates of each sublattice, the system of coupled LLB
equations \eqref{eq:LLBferrimagFinalexpression1} is linearized  by assuming
small deviations from equilibrium,
$m_{\tiem{TM(RE)}}=m_{e,\tiem{TM(RE)}}+\delta m_{\tiem{TM(RE)}}$ and $\delta
\mathbf{m}=(\delta m_\tiem{TM}, \delta m_\tiem{RE})$.  This gives
a characteristic matrix, $\mathcal{A}_{\|}$, which drives the dynamics of the
linearized equation $\partial (\delta \mathbf{m}) / \partial
t =\mathcal{A}_{\|}\delta \mathbf{m}$.  The matrix $\mathcal{A}_{\|}$ reads
\begin{equation}
\label{matrix:matrixFerriLongitudinal}
\mathcal{A}_{\|}=\left(\begin{array}{ccc}
 -\gamma_{\tiem{TM}}\alpha_{\|}^{\tiem{TM}}/\Lambda_{\tiem{TT}}
 \enskip   &
 \gamma_{\tiem{TM}} \alpha_{\|}^{\tiem{TM}}J_{0,\tiem{TR}}/\mu_{\tiem{T}}
   \\

 \gamma_{\tiem{RE}} \alpha_{\|}^{\tiem{RE}}
 J_{0,\tiem{RT}}/\mu_{\tiem{R}} \enskip
  & -\gamma_{\tiem{RE}}\alpha_{\|}^{\tiem{RE}}/\Lambda_{\tiem{RR}}
 \end{array}\right)
= \left(\begin{array}{ccc}
    \Gamma_{\tiem{TT}}   & \Gamma_{\tiem{TR}} \\
    \Gamma_{\tiem{RT}}   & \Gamma_{\tiem{RR}}
 \end{array}\right)
\end{equation}
It is important to note that the matrix elements in equation
\eqref{matrix:matrixFerriLongitudinal} are temperature dependent.  The general
solution of the characteristic equation, $|\mathcal{A}_{\|}-\Gamma^{\pm}
\mathcal{I}| =0$, gives two different eigenvalues, $\Gamma^{\pm}=1/\tau_{\pm}$,
corresponding to the eigenvectors $\mathbf{v}_{\pm}=(\Gamma_{\tiem{TR}},
-(\Gamma_{\tiem{TT}}+\Gamma^{\pm}))$. In ferromagnets, relaxation can usually
be described well by only one relaxation rate, at least in the linear regime.
By contrast, in two sublattice ferrimagnets, a combination of the two
characteristic relaxation rates $\Gamma^{\pm}$ describes the magnetization
relaxation of each sublattice with a weighting determined by the eigenvectors.
This means that one cannot describe the relaxation of a two-component system
with as single exponential decay function except in some limits that we will
consider next.

Figure \ref{fig:Fig1} (b) shows the temperature dependence of the matrix
elements Eq.~(\ref{matrix:matrixFerriLongitudinal}),  for
a Gd$_{25}$(FeCo)$_{75}$ alloy. At temperatures far from $T_c$, $T\leq
(3/4)T_c$, the relaxation rates fulfill the conditions $\Gamma_{\tiem{TT}} \gg
\Gamma_{\tiem{RR}} > \Gamma_{\tiem{RT(TR)}}$. Therefore the motion of each
sublattice is approximately defined by the corresponding diagonal element,
\emph{i.e.} $\partial \delta m_{\tiem{TM(RE)}}/\partial
t=-\Gamma_{\tiem{TT(RR)}}\delta m_{\tiem{TM(RE)}}$, or equivalently, each
eigenvector is associated with one sublattice. The relaxation rates
($\tau^{-1}$) are $-\gamma_{\nu}\alpha_{\|}^{\nu}/\Lambda_{\nu\nu}$, ($\nu=$RE,
TM). It has been shown that  $\gamma_{\tiem{RE}}\alpha_{\|}^{\tiem{RE}}\approx
\gamma_{\tiem{TM}}\alpha_{\|}^{\tiem{TM}}$, \cite{AtxitiaPRB2012}  thus  the
leading contribution to the temperature dependence of the demagnetization times
comes from the parameters $\Lambda_{\nu\nu}$ which can be regarded as an
effective longitudinal susceptibility in analogy with the pure ferromagnetic
case. This parameter in turn depends on the actual longitudinal
susceptibilities as
\begin{figure}[tbc!]
\begin{center}
\includegraphics[width=0.48\textwidth]{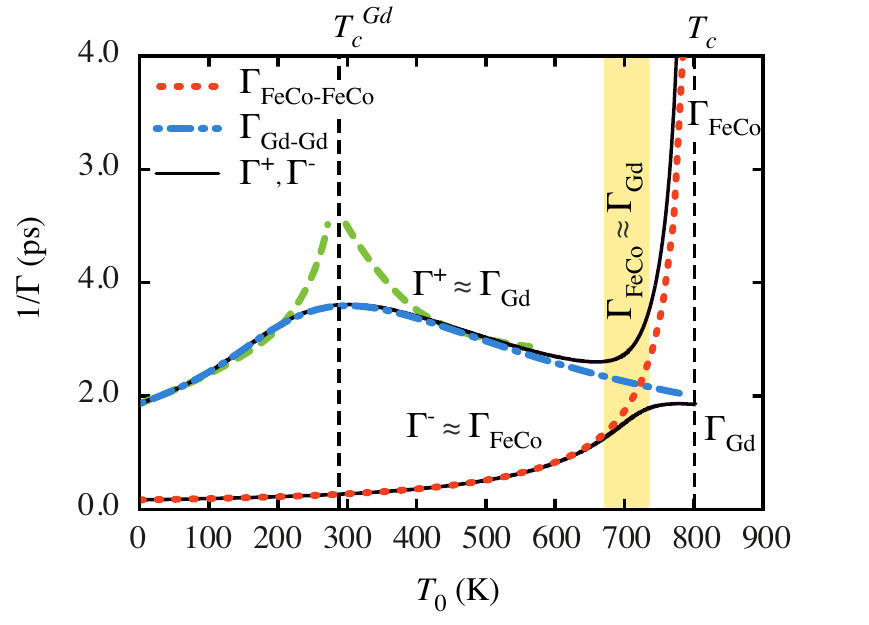}
\end{center}
\caption{Characteristic relaxation times in GdFeCo alloy as a function of temperature. At relatively low temperatures  $\Gamma^{+}\approx \Gamma_{\mathrm{Gd}}$ and $\Gamma^{-}\approx \Gamma_{\mathrm{FeCo}}$.
The Gd relaxation time presents a maximum at $T_c^{\mathrm{Gd}}$ caused by the slowing down of the Gd fluctuations related to Gd-Gd interactions. The yellow shaded area corresponds to a mixed relaxation times  and both sublattices relax similarly.
Close to $T_c$, $\Gamma_{\mathrm{Gd}}\gg \Gamma_{\mathrm{FeCo}}$, and  Gd sublattice magnetization
relaxes faster.}
\label{fig:Figure3}
\end{figure}

\begin{figure}[tbc!]
\begin{centering}
\includegraphics[scale=1]{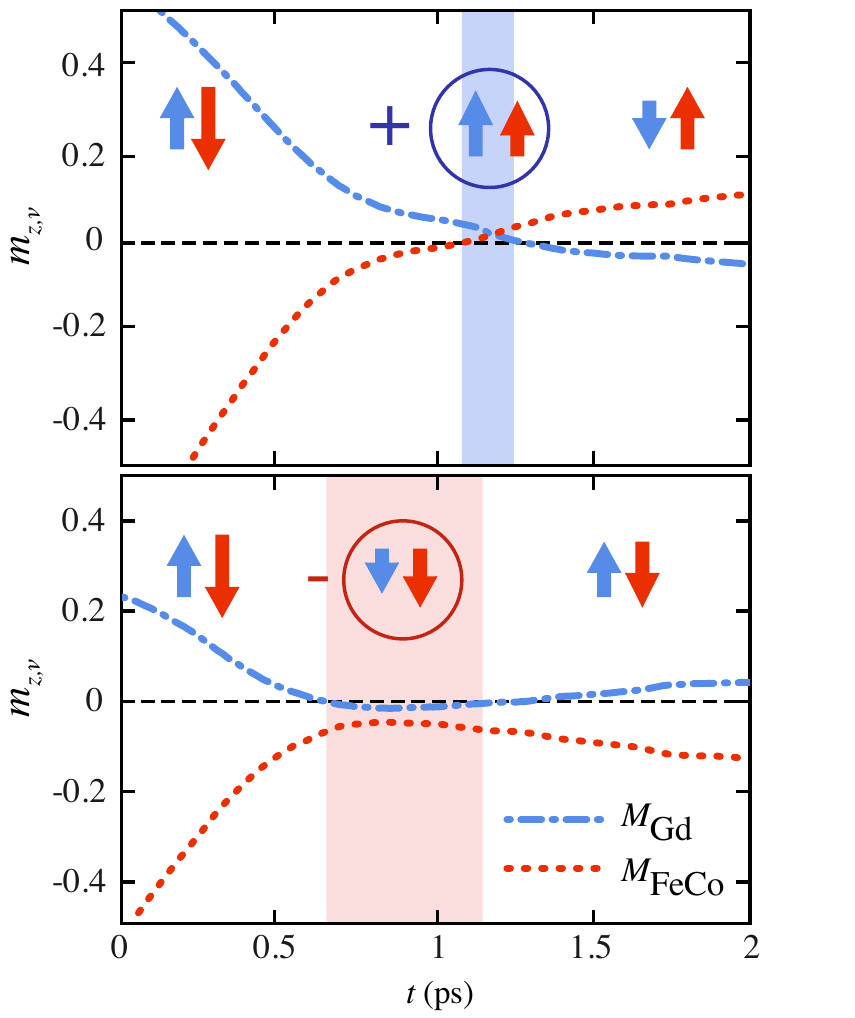}
\end{centering}
\caption{(Color online)  (up) A positive TFLS means that the FeCo sublattice
  reverses first after the application of a laser heat pulse, (down) and a negative TFLS that the Gd sublattice reverses. }
\label{fig:tfls}
\end{figure}

\begin{figure*}[tbc!]
\begin{centering}
\includegraphics[]{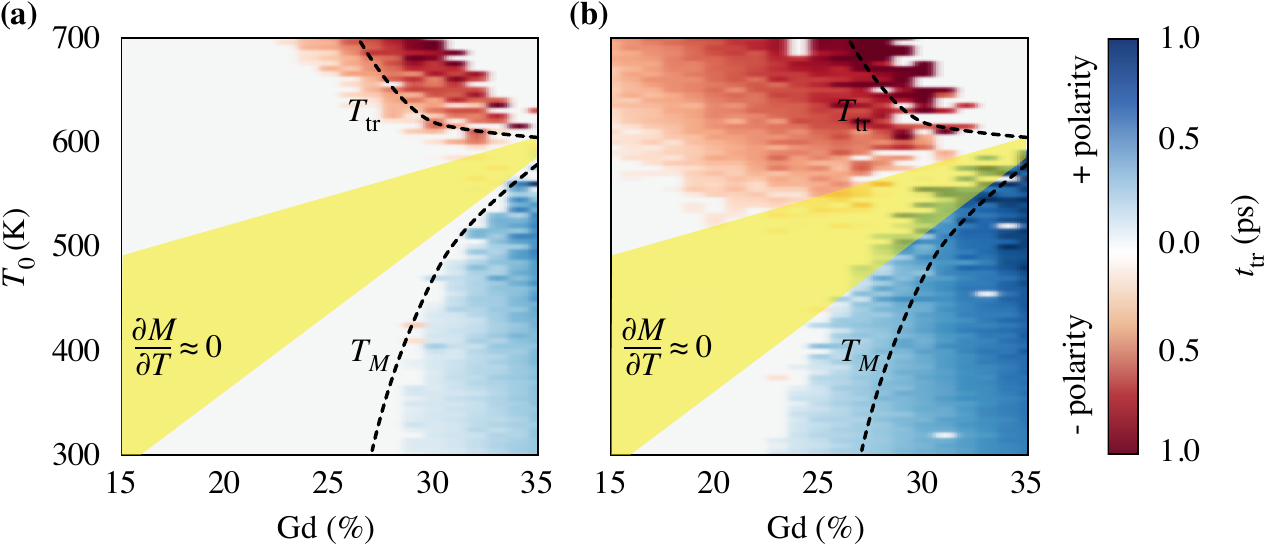}
\end{centering}
\caption{(Color online)  (a) and
  (b) phase diagrams of the TFLS duration and polarity as a function of  the
  initial temperature $T_0$ and Gd concentration for a heat step pulse of
  $600$ fs  and $T_{\mathrm{max}}=1200$ K (a) and $T_{\mathrm{max}}=1500$ K (b)
  gained by computer atomistic simulations. A positive TFLS means that the FeCo
  sublattice reverse first, and a negative TFLS that Gd sublattice does.  The
  dashed line in (a) and (b) corresponds to the transition temperature
  $T_{\mathrm{tr}}$ and $T_{\mathrm{M}}$.  For low excitation heat pulses
($T_{\mathrm{max}}$=1200 K in (b)) these characteristic temperatures define
well the regions where the TFLS can be encountered and its polarity. }
\label{fig:phasediagram}
\end{figure*}
\begin{equation}
  \tau_{\|}^{\nu}=\frac{\Lambda_{\nu\nu}(T)}{\gamma_{\nu}\alpha^{\nu}_{\|}}=
 \frac{1}{\gamma_{\nu}\alpha^{\nu}_{\|}}
 \frac{\widetilde{\chi}^{\nu}_{\|}(T)}{1+\frac{J_{0,\nu\kappa}}{\mu_{\nu}}\widetilde{\chi}^{\kappa}_{\|}(T)}
.
\label{eq:relaxationtime}
\end{equation}

Equation \eqref{eq:relaxationtime} allows us to analyze the differences between
both magnetic sublattices for $T_0\leq (3/4)T_c$. For  RE concentrations
$(x=0.2-0.3)$ where the TFLS has been experimentally observed, the mean field
exchange interaction between TM and RE spins can be neglected,
$J_{0,\tiem{TR}}=x z J_{\tiem{TR}}$ is small given that $J_{0,\tiem{TT}} \gg
J_{0,\tiem{TR}}$.  This leads to the approximate form of
Eq.~\eqref{eq:relaxationtime} as $\gamma_{\tiem{TM}}\alpha^{\tiem{TM}}_{\|}
\tau_{\|}^{\tiem{TM}} (T)= \widetilde{\chi}^{\tiem{TM}}_{\|}(T)\propto
\mu_{\tiem{TM}}/ J_{0,\tiem{TT}}$. This dependence reflects the fact that the TM
relaxation is mainly determined by the exchange interaction between TM
neighbors because each atomic TM moment is almost totally surrounded by other
TM moments ($1-x=0.7-8$).  The RE relaxation is determined by the
antiferromagnetic exchange coupling with the TM neighbors $J_{0,\tiem{RR}}=x
z J_{\tiem{RR}}$ which is small due to the small probability that any RE
moments interact with a RE nearest neighbor at low RE concentrations.  At low
temperatures and rare earth concentrations, one can consider the limiting case
of a dilute paramagnetic rare earth species within a transition metal bulk.
Using $(J_{0,\tiem{TR}}/\mu_{\tiem{RE}})\widetilde{\chi}_{\|}^{\tiem{TM}} \ll
1$, from Eq. \eqref{eq:relaxationtime} one obtains that that
$\gamma_{\tiem{RE}}\alpha^{\tiem{RE}}_{\|}\tau_{\|}^{\tiem{RE}}=
\widetilde{\chi}^{\tiem{RE}}_{\|}(T)$, which explains the local maximum in
$\Gamma_{\mathrm{Gd}}$  at $T_c^{\mathrm{Gd}}=291$ K and which is related to
the slowing down due to ferromagnetic Gd-Gd interactions [dashed lines in Fig.
\ref{fig:Figure3}].  In this temperature region, the relevant magnetic
parameter defining the RE magnetization relaxation is
$\gamma_{\tiem{RE}}\alpha^{\tiem{RE}}_{\|}\tau_{\|}^{\tiem{RE}}=
\mu_{\tiem{RE}}/ |J_{0,\tiem{RT}}|$ and we obtain:
\begin{equation}
  \frac{\tau^{\tiem{RE}}_{\|}}{\tau^{\tiem{TM}}_{\|}}\sim
  \frac{\mu_{\tiem{RE}}|J_{0,\tiem{TT}}|}{\mu_{\tiem{TM}}|J_{0,\tiem{RT}}|} \gg 1.
\end{equation}
As temperature increases, the off diagonal terms in matrix equation
$ |\mathcal{A}_{\|}-\Gamma^{\pm} \mathcal{I}| =0$ start to play an increasing
role. It is no longer possible to attribute one eigenvector to one sublattice
but instead there are mixed contributions from both sublattices. This means
that both the TM and RE sublattices relax with a similar mix of the two
characteristic times, $\Gamma^{\pm}$  [grey zone in Fig. \ref{fig:Figure3}].
Both sublattices are magnetically equivalent in terms of relaxation rates
around  some temperature $T_{\text{tr}}$, where $\Gamma_{\tiem{TT}} \approx
\Gamma_{\tiem{RR}}$ holds.  This finding clearly shows that the simple ratio
between atomic magnetic moments $\mu_{\tiem{TM}}/\mu_{\tiem{RE}}$ is no longer
a sufficiently good estimate of the non-equivalence of the sublattice
relaxation times.

Above $T_{\text{tr}}$ and approaching $T_c$, the TM relaxation time becomes
longer than that of the RE. The TM is undergoing critical slowing down towards $T_c$
\cite{RushbrookeBook} due to the divergence of
$\widetilde{\chi}^{\tiem{TM}}_{\|}$ [see Fig. \ref{fig:Figure3}]. The RE
relaxation time is only slightly temperature dependent even approaching $T_c$
as $\gamma_{\tiem{RE}}\alpha^{\tiem{RE}}_{\|}\tau_{\|}^{\tiem{RE}}= \mu_{\tiem{RE}}/
|J_{0,\tiem{RT}}|\left(\widetilde{\chi}^{\tiem{TM}}_{\|}
/\widetilde{\chi}_{\|}^{\tiem{RE}}\right)$, where the ratio between partial
susceptibilities does not strongly depend on temperature.
\cite{AtxitiaPRB2012LLBferrim}

 We conclude that $T_{\text{tr}}$ separates two \emph{non-equivalence} regimes
 for the element specific magnetization dynamics. Though this result is
 strictly applicable only in the linear regime (small deviations from
 equilibrium), we show next that it can be extended for the laser heating
 induced TFLS and a similar transition can be observed.

\section{Control of the TFLS polarity}

With our knowledge of these two non-equivalent regimes, we shall demonstrate
that one may selectively excite a certain polarity depending on the composition
of the alloy and the initial temperature. To do so, we use the atomistic spin
model described by the Hamiltonian \eqref{eq:binaryalloyGeneralHamiltonian} and
the set of atomistic LLG equations \eqref{eq:stochasticLLG} to perform
extensive computer simulations of  pump-probe laser heating experiments in
GdFeCo alloys for a range of Gd concentrations, initial temperatures, and laser
fluence. The dynamical behavior of $T_e(t)$ after the application of
a femtosecond laser pulse has been shown to be well reproduced by the
two-temperature model (2TM).\cite{Chen2006} Within this model the electron and
phonon systems are described as heat baths with  associated temperatures
$T_e(t)$ and $T_{ph}(t)$.  The laser pulse heats the electron system which
increases its temperature in the timescale of the order of hundreds of
femtoseconds, to around 1000 to 2000 K, afterwards the electron system cools
down by transferring energy to the phonon system in the picosecond time scale
through the electron-phonon coupling.  In our current calculations, this model
is simplified by considering that the effect of the laser is to increase the
electron temperature from the initial temperature $T_0$ to  $T_{\textrm{max}}$
and then to reduce back to $T_0$ (a temperature step with $600$ fs duration in
our model).  We use this simple profile for the electron temperature so that we
may disentangle the magnetization dynamics from the effect of different
temperature profiles which are usually considered in ultrafast pump-probe
experiments.

The results in Fig.~\ref{fig:tfls} clearly show that the TFLS can be
formed in either TM (+) or RE (-) polarities. This depends on the initial
temperature and the Gd concentration.  In Fig.~\ref{fig:phasediagram}(a),(b) we
calculate the polarity and duration of the TFLS for two different pulse powers
with $T_{\textrm{max}}=1200$ and $1500$ K, representing a low and high power
input. The coloring, red or blue, represents the polarity of the state formed
and the intensity of the color represents the duration. The result with a low
power input ($T_{\mathrm{max}}=1200$ K) is that the transition temperature
$T_{\mathrm{tr}}$ lies close to our theoretical prediction, $T_{\mathrm{tr}}$.
This happens because the temperature difference between $T_0$ and
$T_{\mathrm{max}}$ at high temperatures when $T_0 \approx T_c$ is not very
large, and the pulse duration is short so the dynamics are still within the
linear regime. The positive polarity, the one already observed experimentally
(blue area), follows the $T_M$ line, supporting the ferrimagnetic exchange mode
excitation described in detail in Sec. \ref{sec:formation}.

For a higher power input ($T_{\mathrm{max}}=1500$ K),
Fig.~\ref{fig:phasediagram}(b), the TFLS is formed beyond the boundary of our
predicted transition temperature, because the linear approximation is no longer
valid.  Nonetheless, the qualitative shape of these regions is still captured
by our theoretical curves. This confirms that the insights gained from our
theoretical estimations based on the LLB model still are useful in such highly
non-equilibrium situations.

In both high and low power examples, the region of relaxation time equivalence
(marked in yellow) prohibits the formation of a TFLS as we had expected. This
result gives a clear link between the equilibrium temperature measurements that
are likely to be easily measured in experiments and the non-equilibrium
ultrafast laser induced control of the transient ferromagnetic-like state in
ferrimagnets.

The experimental verification of our findings is a real challenge since initial
high temperature measurements using element specific femtosecond resolution
XMCD  would not be straightforward due to the small magnetic signals expected.
However, recent experiments in CoTb alloys have shown that Tb magnetization
demagnetizes faster as the ratio $T_0/T_c$ increases.
\cite{LopezfloresArxiv2013} To demonstrate this they used different Tb
concentrations rather than changing the initial temperature. This result agrees
with our findings. In addition, recent experiments
\cite{GravesNatureMaterials2013} measured the spatial distribution of FeCo and
Gd atoms in a heterogeneous GdFeCo alloy, showing significant clustering of the
RE and TM species into regions with large correlation lengths. Using time
resolved XMCD that probed nanoscale spin dynamics they found that, for some
Gd-rich regions (equivalent to our high Gd concentration alloys), the transient
ferromagnetic like state has Gd polarity. 
Graves \emph{et al.} \cite{GravesNatureMaterials2013}
associated this behavior with a non-local magnetization transfer between Gd-rich and FeCo-rich
regions by superdiffusive spin transport.  This observation also provides at
least indirect support for the formation of the RE- driven TFLS polarity
predicted here. Moreover, the laser heating in their work completely demagnetizes the sample, \emph{i.e.} a high laser fluence scenario and we know from Fig.~\ref{fig:phasediagram} that with a higher fluence the negative polarity state becomes more prevalent. A direct comparison between Fig.~\ref{fig:phasediagram} and these experiments is not possible due to the macroscopic clustering found by Graves \emph{et al.} \cite{GravesNatureMaterials2013} which will cause a distribution of effective Curie temperatures in the sample. Also, the sensitively of the results to the Curie temperatures and compensation temperatures mean that a more detailed knowledge of the magnetic properties of their sample would be needed for a direct comparison. Our results however, show that non-local angular momentum transfer is not necessary to interpret their experimental findings if one applies our theory.

\section{Conclusions}

 In summary, we have shown that the femtosecond laser induced transient
 ferromagnetic-like state (TFLS) polarity in rare earth-transition metal
 (RE-TM) ferrimagnets can be controlled by tuning the initial temperature and
 Gd impurity concentration. We have demonstrated that a non-equivalent heating
 efficiency of the two sublattices is a key criterion for the formation of the
 TFLS. To show this, we have studied in detail the RE-TM ferrimagnetic GdFeCo
 alloy and shown that due to the different nature of the spin fluctuations the
 magnetic response of each sublattice depends distinctly on the temperature.
 Microscopically, the magnetization dynamics of the TM sublattice is dominated
 by ferromagnetic fluctuations that slow down its dynamics approaching $T_c$.
 The dynamics of the rare-earth impurities are dominated by antiferromagnetic
 fluctuations that make its dynamics almost temperature independent. Therefore,
 our results predict that, at high temperatures, the conventionally slow
 relaxing rare earth magnetic impurities may become faster than the transition
 metal magnetization. Our results highlight the requirement for the two
 sublattices to have different relaxation times in order to form a TFLS which
 can lead to switching in the positive polarity. In the region of equivalence
 in the relaxation rate of the two sublattices, no TFLS can be formed and
 therefore switching is also prohibited.  These results could have important
 implications for the design of future all-optical THz devices since we have
 also provided a simple picture of this non-equivalence in terms of measurable
 quantities, $M(T)$ and $\partial M/\partial T$, with the further advantage
 that it can be generalized to other multi-sublattice alloys.

\section{Acknowledgement}
This work was supported by  the Spanish Ministry of Science and Innovation
under the grant FIS2010-20979-C02-02 and by the European Community's Seventh
Framework Programme (FP7/2007-2013) under grant agreement No. 281043,
FEMTOSPIN.  U.A. gratefully acknowledges support from  Basque Country
Government under "Programa Posdoctoral de perfeccionamiento de doctores del
DEUI del Gobierno Vasco".


\end{document}